\documentclass[10pt,twocolumn,final]{IEEEtran}
\usepackage[dvips,final]{graphicx}
\usepackage{latexsym}
\usepackage{amssymb}
\usepackage[cmex10]{amsmath}
\usepackage{amsthm}
\usepackage{bm}
\usepackage{bbm}
\usepackage{array}
\usepackage{balance}
\usepackage[caption=false,font=footnotesize]{subfig}
\usepackage{cite}
\usepackage{algorithmic}
\usepackage{algorithm}
\usepackage{xcolor}
\usepackage{multirow}
\usepackage{pgfplots} 
\usepackage{color}
\usepackage{enumitem}
\usepackage{subfiles}
\usepackage{commath}
\usepackage{nicefrac}
\usepackage[colorinlistoftodos,textwidth=\marginparwidth]{todonotes}
\usepackage{wrapfig}
\usepackage{multirow}
\usepackage{tabu}
\usepackage{subfig}

\allowdisplaybreaks

\pgfplotsset{compat=newest}
\pgfplotsset{plot coordinates/math parser=false}

\title{Tile-Based Joint Caching and Delivery of $360^o$ Videos in Heterogeneous Networks}

\author{Pantelis Maniotis,
	    Eirina Bourtsoulatze, \IEEEmembership{Member,~IEEE}
	    Nikolaos~Thomos, \IEEEmembership{Senior~Member,~IEEE}%
\thanks{P. Maniotis and N. Thomos are with the School of Computer Science and Electronic Engineering, University of Essex, Colchester, United Kingdom (e-mail: \{p.maniotis, nthomos\}@essex.ac.uk). Eirina Bourtsoulatze is with University College London, London, United Kingdom (email: e.bourtsoulatze@ucl.ac.uk). This work has been partially funded by the European Union Horizon 2020 research and innovation programme under the Marie Sklodowska-Curie grant agreement No. 750254.}
}

\begin{document}
\maketitle

\begin{abstract}
The recent surge of applications involving the use of $360^o$ video challenges mobile networks infrastructure, as $360^o$ video files are of significant size, and current delivery and edge caching architectures are unable to guarantee their timely delivery. In this paper, we investigate the problem of joint collaborative content-aware caching and delivery of $360^o$ videos in a video on demand setting. The proposed scheme takes advantage of $360^o$ video encoding in multiple tiles and layers to make fine-grained decisions regarding which tiles to cache in each Small Base Station (SBS), and where to deliver them from to the end users, as users may reside in the coverage area of multiple SBSs. This permits to cache the most popular tiles in the SBSs, while the remaining tiles may be obtained through the backhaul. In addition, we explicitly consider the time delivery constraints to ensure continuous video playback. To reduce the computational complexity of the optimization problem, we simplify it by introducing a fairness constraint. This allows us to split the original problem into subproblems corresponding to Groups of Pictures (GoP). 
Each of the subproblems is then solved with the method of Lagrange partial relaxation. Finally, we evaluate the performance of the proposed method for various system parameters and compare it with schemes that do not consider $360^o$ video encoding into multiple tiles and quality layers, as well as with two variants of the proposed method one that considers layered encoding and SBSs collaboration and another that uses tiles encoding but with no SBSs collaboration. The results  showcase the benefits coming from caching and delivery decisions on per tile basis and the importance of exploiting SBSs collaboration.
\end{abstract}


\section{Introduction}

Nowadays we are witnessing an enormous increase in the traffic of $360^o$ visual content originating from Virtual Reality (VR) and Augmented Reality (AR) applications. This growth is fuelled by the proliferation of devices that can display VR content, e.g. smartphones, tablets, Head Mounted Displays (HMDs) as well as services that offer users immersive multimedia experience such as online gaming. Plethora of VR content files and $360^o$ videos are available for downloading on social media platforms such as YouTube and Facebook.

Each video frame in a $360^o$ video encodes a $360^o$ field of view.  At any given time, a user views a portion of the scene, known as \textit{viewport}. The displayed part of the scene may change according to the movement of the user's headset, thus, offering the user an immersive experience. This, however, comes at the cost of very high delivery bandwidth requirements, as the bitrate required to support a high Quality of Experience (QoE) is much larger than that for a regular video.  The technical challenges related to $360^o$ video delivery can be better understood through the following example. A viewport typically covers $120^o$ of the overall scene and can have a resolution of up to 4K ($3840 \times 2160$) \cite{Viewport-adaptive}, which means that the resolution of the whole $360^o$ video scene can be as high as 12K ($11520 \times 6480$). Assuming that a $360^o$ video requires a frame rate of 60 frames per second (fps) to prevent user's dizziness, the  bit-rate required to deliver a high quality $360^o$ video may exceed 100Mbps. Under these bandwidth requirements, the delivery of $360^o$ video to wireless users in a cellular network given the strict delivery deadlines imposed by VR and AR applications, becomes a very challenging problem. 

To facilitate the delivery of massive video content in cellular networks, mobile network operators may enable edge caching \cite{Edge_caching_1,Edge_caching_2,Edge_caching_3}. In edge caching systems, popular content is stored close to the end-users by being placed in the cache of the SBSs. As a result, when a content request is made, it can be served locally from an SBS cache which holds a copy of the requested content. This limits the usage of the pricey backhaul links as well as reduces the latency  through the use of short range communication, which improves the users' QoE.

Despite the existence of edge caching solutions for video \cite{Joint_caching_placement_User_association,Collaborative_1,Collaborative_2,Collaborative_3,Layered_Poularakis,PoularakisTMC19}, they are not readily deployable for the caching of $360^o$ video files. This is due to the fact that the consumption pattern of $360^o$ video content is significantly different from that of the single view video. Specifically, $360^o$ video frames encode the full $360^o$ scene and require significant storage resources; however, only a small portion of the available scene, i.e., a viewport, is viewed by the user at any given time. This creates an additional degree of freedom in deciding which parts of the video file should be cached in the SBSs, as users are only interested in displaying a part of the scene and the delivery of the entire video may be unnecessary as parts of it will never be displayed. Without explicit consideration of these aspects of $360^o$ video content and given the high-rate low-latency constraints for the delivery of $360^o$ video, the growing bandwidth and caching requirements will put more pressure on mobile network operators' infrastructure, who will struggle to accommodate users' (often diverse) demands for $360^o$ video content. Clearly, there is a need for novel methods to optimize the use of the available bandwidth and storage resources in order to support $360^o$ video delivery and maintain high the users' QoE. 
 
\begin{figure}
    \begin{center}
    \subfloat[Encoding of $360^o$ video in two layers and multiple tiles. ]{{\includegraphics[width=9cm]{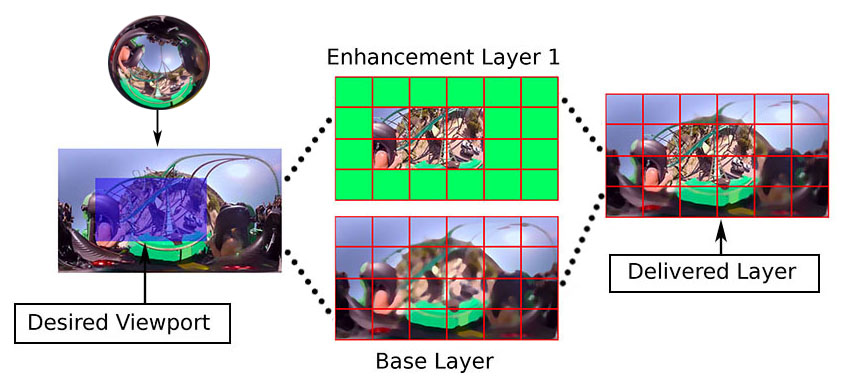} }}
    \qquad
    \subfloat[Users requesting different viewports that overlap.]{{\includegraphics[width=9cm]{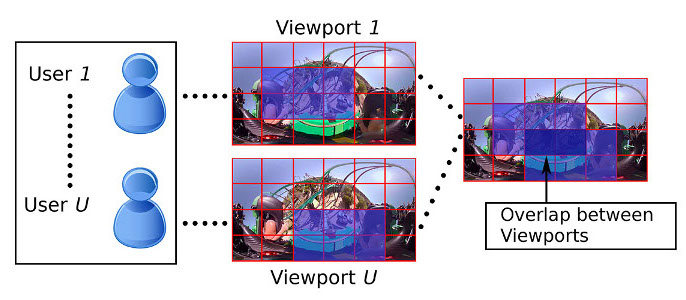} }}
    \end{center}
    \caption{Users request various viewports encoded in multiple layers and tiles, which overlap resulting in different popularity per tile and layer.}
    \label{fig:example}
\end{figure}

In this work, we propose a content-aware $360^o$ video caching and delivery scheme for wireless cellular networks in the context of video on demand applications. Our aim is to maximize the quality of the video delivered to the client population while respecting the delivery deadlines of the $360^o$ video content. To enable the design of efficient joint caching and delivery scheme, we exploit advanced coding tools of video coding standards \cite{Advanced_Comparisson}. Specifically, we leverage the benefits of encoding a $360^o$ video into a number of independent tiles and quality layers. The above video encoding allows us to deal with the diverse demands of the users in terms of the viewports they are interested in watching for each $360^o$ video, in a fine-grained way as illustrated in Fig. \ref{fig:example}.  Furthermore, encoding of $360^o$ video content into tiles and quality layers creates a flexible structure and enables collaborative caching of the various video tiles. This to some extent resembles coded caching schemes \cite{Rel_Work_Collaborative_1} where different parts of the video file can be stored in various SBSs. Through tile encoding, SBSs can aggregately cache only parts of the video content that are important to the users without requiring caches of higher capacity. 


Given the above encoding of $360^o$ videos into tiles and quality layers, we formulate an optimization problem which seeks the joint caching and delivery policy that maximizes the cumulative video distortion reduction. To determine the optimal caching and delivery policy we take into account the physical limitations of the network, i.e., bandwidth and latency, as well as the opportunities for collaboration between the SBSs when users reside in the coverage area of more than one SBSs. To address the fact that content requests cannot be satisfied instantly, we consider that, while users may tolerate a very small startup delay for video streaming applications, an increase of it beyond 2 seconds may cause users to abandon the video, thus resulting in reduction in the perceived QoE \cite{Startup_Delay_Behaviour}. This is captured through the initial playback delay constraint. 

In order to deal with the high complexity of the original problem, we impose a fairness constraint which allows us to decompose the original problem into a set of subproblems. Each subproblem seeks for the optimal caching and delivery policy per one Group of Pictures (GoP). By solving each one of the subproblems and combining their solutions, we obtain the solution of the original problem. The solution of each subproblem is obtained by Lagrangian relaxation and the subgradient algorithm. The Lagrangian subproblem consists of two optimization problems that can be solved independently. The first one is termed as the caching component as it involves only the caching variables. The caching component consists of a set of independent 0-1 knapsack problems which are solved by using Dynamic Programming algorithms \cite{Knapsack_Problems} and determine which tiles should be cached in each SBSs, given a number of constraints described in Section  \ref{sec:prob_formulation}. The second problem is the routing component, as it involves only the routing variables, and can be reduced to the multidimensional multiple choice knapsack problem\cite{Akbara06}.  The routing component of the problem aims at determining where data, that corresponds to tiles, should be fetched from in order to satisfy user requests. The Lagrangian multipliers are updated using the subgradient algorithm which couples again the caching and the routing component.  
Finally, we evaluate and compare the performance of our solution with baseline schemes that do not exploit one or more of the following components: \textit{(i)} tile encoding, \textit{(ii)} layered encoding, \textit{(iii)} SBSs collaboration.
The results illustrate the advantages of the proposed method compared to its counterparts, which do not consider tile encoding and/or cache collaboration. To the best of our knowledge, this is the first work that considers collaborative caching on per-tile basis for $360^o$ videos. An early version of this work has been presented in \cite{ManiotisMMSP19}.


We would like to note that our method is generic and applicable to traditional videos as well. However, differently from the $360^o$ video case where exploiting coding in tiles can improve the performance of the system significantly as we show in this paper, tile encoding will not bring significant gains for traditional videos. The performance gains for $360^o$ videos are attributed the existence of multiple viewports, which results into non-uniform tiles popularity with some tiles being more important than others. This does not happen in the traditional video case, as users request the entire video, and thus tiles popularity is uniform. Therefore, for traditional videos some small gains might be observed only if the decoding delays are such that the delivery of the entire video is not possible, but it is still sufficient to deliver a number of tiles because of their smaller size compared to the entire video. This affects only a limited number of videos which are on the verge of being cached in an SBS because of their relatively low popularity.

In summary, the contributions of our work can be summarized as follows:
\begin{itemize}[leftmargin=*]

\item The introduction of a novel collaborative content-aware caching and delivery scheme that takes advantage of the coding of $360^o$ video in multiple tiles and layers and permits to make fine-grain decisions regarding where to cache and from where to deliver the video tiles to the end users.


    

\item The thorough evaluation of the proposed scheme for various system settings and its comparison with several schemes that do not exploit $360^o$ video encoding in tiles to show the gains coming from the cache and delivery decisions made per tile and layer basis.

\end{itemize}

The rest of this paper is organized as follows. In Section \ref{sec:rel_work}, we give an overview of work related to the caching problem and the $360^o$ video delivery. In Section \ref{sec:sys_setup}, we describe our system setup. Afterwards, we provide the problem formulation in Section \ref{sec:prob_formulation}. In Section \ref{sec:dist_alg}, we propose a reduced complexity algorithm to solve our optimization problem. In Section \ref{sec:sim_results}, we extensively evaluate the performance of the proposed scheme and compare it with other methods. Finally, we draw conclusions in Section \ref{sec:concl}.

\section{Related Work}
\label{sec:rel_work}
In this section, we briefly review the literature related to edge caching, and tile-based $360^o$ video streaming.

Edge caching has been proposed as an efficient way to improve the performance of cellular networks in terms of the observed delivery delay \cite{Joint_caching_placement_User_association,Collaborative_1,Collaborative_2,Collaborative_3,Layered_Poularakis,PoularakisTMC19},  energy consumption \cite{Green_perspective}, and operating cost \cite{ZhangTVT17}. These benefits of edge caching are attributed to the fact that only a small portion of available content accounts for most of the traffic load \cite{Charact_of_YouTube}. Hence, by caching the most popular content at the SBSs, significant resources are saved.  In \cite{Rel_Work_Edge_Caching_2}, it is explored how to capture content popularity dynamics among the various caches by using at each SBS the information about its neighboring SBSs. The impact of content request patterns and users' behavior on edge caching-assisted mobile video streaming of real-life data sets is studied in \cite{Rel_Work_Edge_Caching_3}.

Users' QoE can be improved by taking into account users' association with more than one SBSs \cite{Joint_caching_placement_User_association} and by allowing collaboration between the SBSs \cite{Collaborative_1,Collaborative_2,Collaborative_3}. Collaboration opportunities emerge due to the densification of SBSs in 5G and beyond mobile networks. As a result of SBSs network densification, users may reside in the coverage area of multiple SBSs. When a content request arrives at an SBS which does not hold a copy of this content in its cache, the request can be forwarded to another (possibly neighboring) SBS within the user's communication range, instead of being redirected to distant back-end servers through backhaul links. Hence, content retrieval is accelerated, and the usage of expensive backhaul links is avoided. The benefits of collaborative edge caching have been studied in \cite{Rel_Work_Collaborative_1,Rel_Work_Collaborative_2}, 
where it is shown that the overall cache hit ratio is increased when SBSs decide collaboratively which video files to cache. In \cite{Rel_Work_Collaborative_1}, collaborative edge caching is investigated from an economic point of view. This work explores how to maximize the profit earned when a video request is served by an SBS in two cases:  a) when the video file is entirely cached in only a single cache, and b) when the video is encoded by means of network coding and the resulting coded data is split into a number of segments that can be stored separately in various SBSs. The latter approach allows the problem to be solved linearly. The impact of collaborative caching for video streaming systems in mobile networks is studied in \cite{Rel_Work_Collaborative_2}. From the evaluation, it becomes obvious that collaboration between caches enhances the performance of the overall edge caching system in terms of both the cache hit ratio and the QoE perceived by the users.


Caching systems can be further improved and offer higher QoE to the end users by exploiting video encoding into multiple layers (layered video)  which can be done by scalable coders such as \cite{SHVC_overview}. Specifically, in \cite{Layered_Poularakis,PoularakisTMC19}, the flexibility in caching decisions coming from encoding in multiple layers is used to minimize the delivery delay at the end-users by allowing the video layers to be cached at different SBSs of a network operator. To solve the problem a pseudo-polynomial approximation algorithm is proposed. For a similar setting, another heuristic algorithm is presented in \cite{ZhanCommLetters17}. The work in \cite{PoularakisTMC19} also examines the case of several network operators collaboratively optimizing their caching strategies. In this setting, SBSs' cache space is split into two parts: one allocated for contents requested by users of the operator who owns the cache and a second which is given for caching video layers of content requested by users who belong to other operators. In \cite{Rel_Work_Edge_Caching_1}, the caching cost, the available cache capacity at the SBSs, and the various social traits of the mobile users are considered to decide which video layers to cache in each SBS.

The delivery of $360^o$ video has been studied in \cite{CheungICIP17}, where the objective is to design multiple VR streams of different view ranges and find the optimal head angle-to-stream mapping under the bandwidth and storage constraints. This is done to avoid drifting as coded VR streams tend to overlap in view ranges. A viewport-adaptive navigable $360^o$ video communication system has been presented in \cite{Viewport-adaptive}. This work suggests that sending full $360^o$ video files to the users from where they can extract the viewport of interest leads to waste of bandwidth. On the other hand, sending only the demanded part of the viewport may lead to unacceptable delays. The solution proposed in \cite{Viewport-adaptive} is a viewport-adaptive DASH-compatible scheme, where video representations are offered not only in multiple rates, but also with different quality of emphasized regions. The impact of the response delay on viewport-adaptive streaming of $360^o$ videos is studied in \cite{Impact_Delays}. The results show that viewport adaptive streaming is only efficient under short adaptation intervals and short client buffering delays.

The encoding of a $360^o$ video file into tiles for the purpose of streaming has been investigated in \cite{Tile_approach_1,Tile_approach_2,Tile_approach_3}. This approach has shown significant performance improvement in streaming systems without the use of edge caching. These streaming systems take into account that users are interested in viewing only a viewport of the $360^o$ video scene, and hence there is no need to deliver the whole scene in high quality. The selection of the tiling scheme is investigated in \cite{Rel_Work_Tile-encoding_2}, along with its impact on the compression and the resulting streaming bitrate. Each  $360^o$ video is encoded in two versions with different resolutions, while each version is further encoded into tiles. Similarly to \cite{Rel_Work_Tile-encoding_2}, in \cite{Perf_measurements} the authors investigate the performance of tile-based video streaming in 4G networks with respect to the coding efficiency and the bandwidth savings. The authors of \cite{Navigation-Aware_Adaptive_Streaming} propose a video streaming scheme that aims to optimize the rate at which each tile is downloaded in the context of DASH streaming. In this work, the goal is to maximize the QoE that users experience during the navigation of the $360^o$ video files. The performance of a DASH-based $360^o$ video streaming system is evaluated in a real-world data set \cite{SunJESTCS19}. The results show that using standard machine learning methods and observing a short window of previous requests is sufficient to predict the next viewport accurately. The performance of DASH based $360^o$ video streaming systems is further augmented when video is encoded in layers and the bandwidth allocation between the layers is decided by a rate allocation algorithm \cite{QianMobiCom18}. A viewport prediction algorithm is used to predict the viewport user consumption pattern.

In \cite{Study_on_QoE}, QoE for streaming of $360^o$ videos is studied in terms of: a) perceptual quality, b) presence, c) acceptability, and d) cyber-sickness. In particular, perceptual quality and the acceptability are measured when users watch a $360^o$ video both with or without a Head-Mounted Display (HMD). Similarly, in \cite{Metric_for_360}, a number of objective quality metrics, including PSNR, weighted to spherically uniform PSNR (WS-PSNR), spherical PSNR without interpolation (S-PSNR-NN), spherical PSNR with interpolation (S-PSNR-I), and PSNR in Crasters Parabolic Projection (CPP-PSNR) are explored for quantifying the quality of $360^o$ video encoding. It is concluded in \cite{Metric_for_360} that traditional PSNR is the most appropriate quality metric for measuring end-to-end distortion because of its low complexity. Although we use MSE in this paper as quality metric, our method is transparent to the chosen distortion metric and can be used in conjunction with any other quality metric, e.g. SSIM.

\section{System Setup}
\label{sec:sys_setup}

Differently from the existing methods in the literature, in this paper we examine jointly the caching and delivery of $360^o$ video. We consider that the caching occurs at the SBSs on a per-tile basis. We further make use of SBSs collaboration to serve most of the requests from local caches. In the following, we introduce the system model and explain the various parts of the considered network architecture. 

\label{sec:sys_model}
\subsubsection{Network}
\label{sec:net_model}
\begin{figure}[t]
    \centering
	{\includegraphics[width = 0.5 \textwidth, height=0.37 \textwidth ]{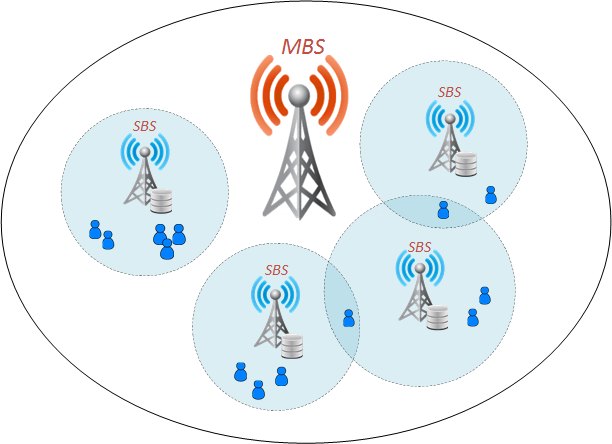}}
	\caption{Mobile network architecture consisting of multiple SBSs and a single MBS. Due to dense placement of SBSs, users can reside in the coverage area of multiple SBSs.}
	\label{fig:Sys_setup}
\end{figure}

We consider a mobile network consisting of a set of $N$ Small-cell Base Stations, as shown in Fig. \ref{fig:Sys_setup}. Let $\mathcal{N}=\{1,\dots\,n,\dots,N\}$ denote the set of SBSs indices. We also assume that the SBSs can communicate with a Macro-cell Base Station (MBS) indexed by ${N+1}$, through which they can retrieve the requested content ($360^o$ video files) from distant servers via backhaul links. For notational convenience, we define the set $\mathcal{N_B}= \mathcal{N} \cup \{{N+1}\}$ which includes the indices of all the SBSs and the MBS.

In the considered network, there are $U=|\mathcal{U}|$ users, where $\mathcal{U}=\{1,\dots, U\}$
is the set of user indices. Each user is associated with one or more SBSs. The association of the users with the SBSs depends on the transmission range of each SBS. The values of the transmission range of the SBSs form the set $\mathcal{R}=\{r_1,\dots\,r_n,\dots,r_N\}$, where $r_n$ is the communication radius of the $n$th SBS. The transmission range of the MBS is denoted by $r_{N+1}$, and is considered to be large enough so that each SBS can establish a connection with the MBS. The overlap in the coverage areas of the SBSs allows users to access multiple SBSs. We introduce the binary variable $\alpha_{nu}\in \{0,1\}$ that equals to $1$ when user $u$ resides in the coverage area of the SBS $n$, and $0$ otherwise. 

Each SBS is equipped with a cache with capacity $C_n \geq 0,\;\forall n \in \mathcal{N}$, where content files can be cached. User's request can be satisfied by any of the associated SBSs, if the requested $360^o$ video file is stored in their cache. Differently, when the file is not cached in any of the SBSs associated with the user, the requested content is fetched from a remote content server through the backhaul link with the help of the MBS. 

We assume that content retrieval is carried out in two phases, namely content placement, and content delivery. In the content placement phase, content is fetched during off-peak hours from distant back-end servers to the caches of the SBSs. In the delivery phase, the cached content is delivered to the users by either the SBSs or through the backhaul link via the MBS, according to the users requests. If the requested content is not cached in the local SBSs, it first has to be fetched via the backhaul, and then it can be delivered to the user. Offline caching can help to avoid network congestion during peak hours, as user requests are diverted from the remote content servers to the local caches (SBSs).

\subsubsection{Video Library}
\label{sec:video_library}
We assume that the video content catalogue contains $V=|\mathcal{V}|$ files, where $\mathcal{V}=\{1,\dots\,v,\dots,V\}$ represents the set of indices of the $360^o$ videos in the catalogue. All the video files are stored at back-end content servers, while SBSs cache only part of the available content catalogue. The caching decisions are taken offline and the files to be cached are moved to the SBSs caches during the content placement phase. 

Each $360^o$ video file is encoded in a set of Groups of Pictures (GoP) $\mathcal{G}$, which in turn are each encoded in a number of independently coded tiles. The tiles of the $v$th $360^o$ video file form the set $\mathcal{T}_v$. For notational convenience, we omit the subscript $v$ in $\mathcal{T}_v$ and use $\mathcal{T}$ for all video files as we assume that all $360^o$ video files are encoded in the same number of tiles. The use of independently encoded tiles is motivated by the fact that users are interested in watching different viewports of the demanded videos. That is, while users may request different video files, they may also be interested in watching different parts of a video scene, i.e., different viewports. As a result, the popularity of tiles depends on both video and viewport popularity. In particular, the overlap between the various viewports shapes the popularity of each tile. Thus, not only the popularity across the viewports of the same video may be different, but also the tiles within the same viewport may have different popularity due to overlap among requested viewports (see Fig. \ref{fig:example}(b)). This in turn affects the optimal caching and routing decisions.

Each tile is further encoded in a set of quality layers $\mathcal{L}$. The most important layer is called base layer. When the base layer is received by a user, it offers a reconstruction of a tile at the lowest available quality. The next layers are known as enhancement layers and contain information that can improve the reconstruction quality of each tile. However, in order to reconstruct a tile at the quality that corresponds to an enhancement layer, all previous enhancement layers including the base layer must be available to the user. Encoding of the $360^o$ video files in layers and tiles offers greater flexibility in deciding which data should be stored in the SBSs caches. Thus, when some viewports are more popular than the others or/and there is significant overlap between some of the viewports, the tiles that form these viewports/overlap regions can be cached at higher quality at the SBSs, if there is sufficient space, while the rest of tiles can be cached in lower quality. 

 When a user requests a viewport of a $360^o$ video file, this request translates into requests for all the tiles of the frame at base layer and all the tiles of the requested viewport at the highest available quality. While the delivery of only the desired viewport would save bandwidth resources, it is not, in general, an optimal strategy. This is due to the fact that, as users navigate through the scene, the actual video consumption pattern may deviate from the expected requests used to determine the joint caching and routing policy. If only the requested viewport was delivered, any deviation in viewing pattern would require a re-transmission of the correct viewport, which in turn would lead to large switching delays and bandwidth waste \cite{Navigation-Aware_Adaptive_Streaming}. By delivering the entire frame at base layer we accommodate for any deviations of the actual viewing pattern of the users from the user requests. Hence, rapid degradation of the perceived QoE is avoided.

\section{Problem formulation}
\label{sec:prob_formulation}

In this paper, we aim at finding the optimal joint caching and delivery policy for tile-encoded $360^o$ layered videos that maximizes the cumulative distortion reduction experienced by the users. To determine the optimal policy we take into account network formation, SBSs capabilities, channel characteristics, users requests, video-specific limitations arising from the encoding of $360^o$ video in tiles and layers, and the requirement for smooth playback. 

Let us introduce the binary variable $y^{nu}_{vglt}\in \{0,1\}$, where $y^{nu}_{vglt}=1$, if the tile $t\in \mathcal{T}$ of the layer $l\in \mathcal{L}$ that belongs to GoP $g\in \mathcal{G}$ of the video $v\in \mathcal{V}$ will be delivered directly from the SBS $n\in \mathcal{N}$ to the user $u\in \mathcal{U}$, and $y^{nu}_{vglt}=0$ otherwise. Similarly, let $y^{(N+1)u}_{vglt}\in \{0,1\}$ be equal to $1$ when a tile is fetched from the backhaul through the MBS, and $0$ otherwise. Therefore, the routing decisions in our system can be described by the vector\footnote{Boldfaced letters correspond to vectors.}:

\begin{equation}
\begin{gathered}
\small
	\mathbf{y} = (y^{nu}_{vglt} \in \{0,1\}:\\ n \in \mathcal{N_B}, u \in \mathcal{U}, v \in \mathcal{V}, g \in \mathcal{G}, l \in \mathcal{L}, t \in \mathcal{T} )
	\label{eq:routing_var}
\end{gathered}
\end{equation}

\noindent We note that requests for tiles are \textit{unsplittable}. This means that each tile request is entirely satisfied either by only one SBS or it has to be fetched through the backhaul. 

Now, let us define the binary decision variable $x^{n}_{vglt}\in \{0,1\}$, which takes the value $1$ when the tile $t\in \mathcal{T}$ of the layer $l\in \mathcal{L}$ that belongs to GoP $g\in \mathcal{G}$ of the video $v\in \mathcal{V}$ is cached at the $n$th SBS, and $0$ otherwise. Hence, the caching decisions for the entire system are described by the vector:

\begin{equation} 
\small
	\mathbf{x} = (x^{n}_{vglt} \in \{0,1\}:n \in \mathcal{N}, v \in \mathcal{V}, g \in \mathcal{G}, l \in \mathcal{L}, t \in \mathcal{T} )
	\label{eq:caching_var}
\end{equation}

Recall that SBSs have limited cache capacity enough only to store a number of tiles from the tile-encoded $360^o$  layered videos.  If $o_{vglt}$  denotes the size of a tile (in \textit{Mbits}), we have:

\begin{equation} 
\small
    \sum_{v \in \mathcal{V}} \sum_{g \in \mathcal{G}} \sum_{l \in \mathcal{L}} \sum_{t \in \mathcal{T}} o_{vglt} x^n_{vglt} \leq C_n, \forall{n \in \mathcal{N}}
	\label{eq:cache_con}
\end{equation}
where $C_n$ the cache capacity (in \textit{Mbits}) of the $n$th SBS. Eq. \eqref{eq:cache_con} is the cache capacity constraint.

Another constraint of the optimization problem arises from the fact that in order to deliver a tile requested by a user from the cache of an SBS, the tile has to be stored in the cache and the user has to reside in the coverage area of the SBS. Hence, it should hold that:

\begin{equation} 
\begin{gathered}
\small
    y^{nu}_{vglt}\leq \alpha_{nu} x^n_{vglt}, \\ \forall n\in \mathcal{N}, u\in \mathcal{U},  v\in \mathcal{V}, g\in \mathcal{G}, l\in \mathcal{L}, t\in \mathcal{T}
 	\label{eq:dev_only_if_cache_con}
\end{gathered}
\end{equation}
Recall that $\alpha_{nu}$ is a binary variable that takes value 1 when user $u$ resides in the coverage area of the $n$th SBS, otherwise its value is 0.

The following constraint ensures that each tile will be received only once by each client:

\begin{equation} 
\small
    \sum_{n\in \mathcal{N_B}}y^{nu}_{vglt}\leq 1, \forall u\in \mathcal{U}, v\in \mathcal{V}, g\in \mathcal{G}, l\in \mathcal{L}, t\in \mathcal{T}
 	\label{eq:unsplittable_con}
\end{equation}

In addition, we have to take into consideration limitations arising from the encoding in multiple tiles and layers. Hence, in order to recover the video in the lowest available quality, only the base layer should be delivered to the user. Differently, in order to achieve the quality that corresponds to the $l$th enhancement layer, the user should receive not only the $l$th enhancement layer, but also the base layer and all the previous enhancement layers. Therefore, we have:

\begin{equation}
\begin{gathered}
\small
    \sum_{n \in \mathcal{N_B}}y^{nu}_{vg(l+1)t}\leq \sum_{n \in \mathcal{N_B}}y^{nu}_{vglt}, \\ \forall   u\in \mathcal{U}, v\in \mathcal{V}, g\in \mathcal{G}, l\in \mathcal{L}, t\in \mathcal{T}
 	\label{eq:layer_con}
\end{gathered}
\end{equation}

The requirement for smooth playback introduces another constraint to our system. The video packets should reach the user within a specific time constraint. When this does not happen, buffer underrun occurs, where the buffer is fed with data at a lower speed than the data is consumed. Let us define $d_{nu}$ as the time needed to transmit a tile $t \in \mathcal{T}$ from the $n$th SBS to the user $u$, and $d_{(N+1) u}$ the corresponding time when the tile is fetched from the backhaul. Obviously, $d_{(N+1) u} > d_{nu}, \; n \in \mathcal{N}$ due to the additional time needed to transmit the requested tile from the backhaul to the MBS. Therefore, it should be:

\begin{equation} 
\begin{gathered}
\small
    \sum_{n \in \mathcal{N_B}} \sum_{g^{'} \in \{1,...,g\}} \sum_{l \in \mathcal{L}} \sum_{t \in \mathcal{T}} o_{vg^{'}lt}  d_{nu}  \alpha_{nu} y^{nu}_{vg^{'}lt} \\  \leq  t_{app}+(g-1)  t_{disp},  \forall u \in \mathcal{U}, v \in \mathcal{V}, g \in \mathcal{G}
 	\label{eq:time_con}
\end{gathered} 	
\end{equation}
where $t_{app}$ and $t_{disp}$ correspond to the playback delay and the time needed to display a GoP, respectively. 

When a user obtains a tile, the distortion observed by the user decreases. Let us denote by $\delta_{vglt}$, the average distortion reduction associated with tile $t\in \mathcal{T}$ of layer $l\in \mathcal{L}$ that belongs to GoP $g\in \mathcal{G}$ of the video $v\in \mathcal{V}$.  The expected achievable cumulative distortion reduction $\Delta$ across the user population is given by:
\begin{equation}
\begin{gathered}
\small
    \Delta=\sum_{u\in \mathcal{U}} \sum_{v\in \mathcal{V}} \sum_{g\in \mathcal{G}}\sum_{l\in \mathcal{L}}\sum_{t\in \mathcal{T}} z^u_{vglt}\delta_{vglt}
 	\label{eq:sum_dis_red}
\end{gathered}
\end{equation}
where the variable $z^u_{vglt}\in [0,1]$ denotes the probability of user $u$ to request the $l$th layer of tile $t$ that belongs to the $g$th GoP of the $v$th video file. This probability can be computed based on the video  and the viewport popularity distributions.

The optimization objective is to maximize the normalized expected cumulative distortion reduction over the client population. Denoting this quantity by $D$, we have:

\begin{equation}
\begin{gathered}
\small
    D=\frac{1}{\Delta} \sum_{n\in \mathcal{N_{B}}} \sum_{u\in \mathcal{U}} \sum_{v\in \mathcal{V}} \sum_{g\in \mathcal{G}}\sum_{l\in \mathcal{L}}\sum_{t\in \mathcal{T}} z^u_{vglt} \delta_{vglt}\alpha_{nu}y^{nu}_{vglt}
 	\label{eq:objective}
\end{gathered}
\end{equation}

By taking into consideration the above constraints, the problem of joint caching and delivery of $360^o$ videos can be formally expressed as:
\begin{equation}
\begin{aligned}
& \mathcal{P}:\underset{\mathbf{x}, \mathbf{y}}{\max} D \\
& \textit{s.t.:}
& & (\ref{eq:routing_var})-(\ref{eq:time_con})
\label{eq:Problem_formulation}
\end{aligned}
\end{equation}

The key notation of our problem is summarized in Table \ref{table:system}.

\begin{table}
\centering
\captionsetup{font=scriptsize}
\caption{Notation}
\begin{tabular}{ |c|m{7cm}| } 
\hline
\textbf{Symbol} & \textbf{Physical Meaning} \\
\hline
\hline
$\mathcal{V}$ & Set of $360^o$ videos  \\ 
\hline
$\mathcal{G}$ & Set of GoPs  \\ 
\hline
$\mathcal{L}$ & Set of quality layers \\ 
\hline
$\mathcal{T}$ & Set of encoded tiles  \\ 
\hline
$\mathcal{N}$ & Set of Small Base Stations  \\ 
\hline
$\mathcal{N_B}$ & Set of SBSs and MBS\\ 
\hline
$\mathcal{U}$ & Set of users  \\ 
\hline
$z^u_{vglt}$ & {Probability that user $u$  requests tile $vglt$}\\
\hline
$\delta{vglt}$ & Distortion reduction from obtaining the tile $vglt$  \\
\hline
$\alpha_{nu}$ & Association of user $u$ with the $n$th SBS \\
 \hline
$o_{vglt}$ & Size of the tile $vglt$ \\ 
\hline
$C_n$ & Cache capacity of the SBS $n$ \\
\hline
$x^n_{vglt}$ & Caching variable for tile $vglt$ at the $n$th SBS\\
\hline
$y^{nu}_{vglt}$ & Routing variable for tile $vglt$ from the $n$th SBS or the MBS to user $u$ \\
 \hline
$t_{app}$ & Playback delay  \\ 
\hline
$t_{disp}$ & Time duration of a GoP \\
\hline
\end{tabular}
\label{table:system}
\end{table}


\section{Distributed Algorithm}
\label{sec:dist_alg}

The optimization problem in Eq. (\ref{eq:Problem_formulation}) is NP-hard. This is because, as we will show later, it can be decomposed into a number of subproblems, where each subproblem consists of two components that are NP-hard. The first component is the caching component that can be translated to a set of 0-1 knapsack problems, while the second component can be translated to the multidimensional multiple choice knapsack problem. Both the 0-1 knapsack problem and the multidimensional multiple choice knapsack problem are known to be NP-hard \cite{Knapsack_Problems}. 

To solve the problem efficiently, we simplify the original problem by introducing a fairness constraint with respect to the cache space allocation and the delivery delay per generation. Specifically, in order to ensure fairness regarding the cache, we limit the cache available for each generation to $\lfloor C_n/G \rfloor$. Similarly, for the delivery delay, we assume that the delay for delivering each generation is $t_{app}/G+t_{disp}$. This allows us to decompose the original problem in (\ref{eq:Problem_formulation}) into $G$ subproblems $\mathcal{P}_1,\dots\,\mathcal{P}_g,\dots,\mathcal{P}_G$, where: 

\begin{align}
& \mathcal{P}_g:\underset{\mathbf{x}_g,\mathbf{y}_g}{\max} D_g \label{eq:Problem_formulation2}\\
& \textit{s.t.:} \nonumber\\
&\mathbf{y}_g = (y^{nu}_{vglt} \in \{0,1\}: n \in \mathcal{N_B}, u \in \mathcal{U}, v \in \mathcal{V}, l \in \mathcal{L}, t \in \mathcal{T} ) \label{eq:routing_var_gop} \\
&\mathbf{x}_g = (x^{n}_{vglt} \in \{0,1\}:n \in \mathcal{N}, v \in \mathcal{V}, g \in \mathcal{G}, l \in \mathcal{L}, t \in \mathcal{T} ) \label{eq:caching_var_gop} \\
& \sum_{v \in \mathcal{V}} \sum_{l \in \mathcal{L}} \sum_{t \in \mathcal{T}} o_{vglt} x^n_{vglt} \leq C_n(g), \forall{n \in \mathcal{N}} \label{eq:cache_con_gop} \\
& \sum_{n \in \mathcal{N_B}} \sum_{l \in \mathcal{L}} \sum_{t \in \mathcal{T}} o_{vglt} d_{nu} \alpha_{nu} y^{nu}_{vglt} \leq t(g), \forall u \in \mathcal{U}, v \in \mathcal{V} \label{eq:delay_con_gop}\\
& y^{nu}_{vglt}\leq \alpha_{nu} x^n_{vglt}, \forall n\in \mathcal{N}, u\in \mathcal{U},  v\in \mathcal{V}, l\in \mathcal{L}, t\in \mathcal{T} \label{eq:dev_only_if_cache_con_gop} \\
& \sum_{n\in \mathcal{N_B}}y^{nu}_{vglt}\leq 1, \forall u\in \mathcal{U}, v\in \mathcal{V},  l\in \mathcal{L}, t\in \mathcal{T} \label{eq:unsplittable_con_gop} \\
& \sum_{n \in \mathcal{N_B}}y^{nu}_{vg(l+1)t}\leq \sum_{n \in \mathcal{N_B}}y^{nu}_{vglt},  \forall   u\in \mathcal{U}, v\in \mathcal{V}, l\in \mathcal{L}, t\in \mathcal{T} \label{eq:layer_con_gop}
\end{align}
where 

\begin{equation}
\begin{gathered}
\small
    D_g=\frac{1}{\Delta} \sum_{n\in \mathcal{N_{B}}} \sum_{u\in \mathcal{U}} \sum_{v\in \mathcal{V}}\sum_{l\in \mathcal{L}}\sum_{t\in \mathcal{T}} z^u_{vglt} \delta_{vglt}\alpha_{nu}y^{nu}_{vglt}
 	\label{eq:objective_gop}
\end{gathered}
\end{equation}


In the above subproblem, $\mathbf{x}_g$ and $\mathbf{y}_g$ correspond to the cache and routing variables for the $g$th GoP.  Eq. (\ref{eq:cache_con_gop}) is the cache constraint of the subproblem $\mathcal{P}_g$. $C_n(g)$ stands for the cache space available for the $g$th GoP of SBS $n$ and is defined as:

\begin{equation}
C_n(g)=\lfloor C_n/G+C^{rem}_n(g-1)\rfloor,
\label{eq:cacheupdate}
\end{equation} 
where $C^{rem}_n(g-1) \geq 0$ corresponds to the amount of cache that has not been filled in with content after solving subproblem $\mathcal{P}_{g-1}$. Similarly, Eq (\ref{eq:delay_con_gop}) is the delay constraint for the subproblem $\mathcal{P}_g$, where $t(g)$ represents the delivery delay of the $g$th GoP and is calculated as:
\begin{equation}
t(g)= t_{app}/G +t_{disp}+t^{rem}(g-1)
\label{eq:delayupdate}
\end{equation} 
with $t^{rem}(g-1) \geq 0$ being the time remaining from the delivery of the previous generation. We would like to note that the above approach is one of the possible ways to impose fairness in cache allocation between GoPs and that the main aim of this paper is to show the advantages arising from the use of tiles and layers in $360^o$ video caching and delivery. The employment of more advanced fairness policies would further improve the performance of the proposed system, however, it would not affect the derived conclusions regarding the benefits of exploiting tiles and layered encoding in caching and delivery systems. To obtain the global policy for routing and caching, i.e., $\textbf{x}$ and $\textbf{y}$, we solve each subproblem sequentially starting from the first generation. After solving the subproblems for all generations, we combine the optimal routing and caching solutions $\textbf{x}_g$ and $\textbf{y}_g$ to obtain the global routing and caching policy. This procedure is summarized in Algorithm \ref{algo:Proposed}.

\begin{algorithm}[t]
\baselineskip=18pt
\caption{Proposed Algorithm}
\label{algo:Proposed}
\begin{algorithmic}[1]
\STATE Decompose $\mathcal{P}$ into $ \mathcal{P}_1,\dots,\mathcal{P}_g,\dots,\mathcal{P}_G$ according to Eq. (\ref{eq:Problem_formulation2})
\STATE Set generation index $g \leftarrow 1$
\STATE Set cache $C_n(1) \leftarrow \lfloor C_n/G\rfloor, \; \forall n \in \mathcal{N}$
\STATE Set delay $t(1) \leftarrow t_{app}/G + t_{disp}$ 
\WHILE {$g \leq G$ }
\STATE Determine optimal delivery and cache policy $\mathbf{x}_g, \mathbf{y}_g$ for $g$th GoP by solving $\mathcal{P}_g$
\STATE Compute  $\forall n \in \mathcal{N}$ \\ 
$C_n^{rem}(g) \leftarrow C_n(g) - \sum\limits_{v \in \mathcal{V}} \sum\limits_{l \in \mathcal{L}} \sum\limits_{t \in \mathcal{T}} o_{vglt} x^n_{vglt}$
\STATE Compute \\
$t^{rem}(g) \leftarrow t(g) - \sum\limits_{n \in \mathcal{N_B}} \sum\limits_{l \in \mathcal{L}} \sum\limits_{t \in \mathcal{T}} o_{vglt} d_{nu} \alpha_{nu} y^{nu}_{vglt}$
\STATE Update cache $\mathcal{C}_n(g+1) \leftarrow \lfloor C_n/G + C_n^{rem}(g)\rfloor $
\STATE Update delay $t(g+1) = t_{app}/G + t_{disp} + t^{rem}(g)$
\STATE $g \leftarrow g+1$
\ENDWHILE
\STATE $\mathbf{x}=\bigcup_{g \in \mathcal{G}}  \mathbf{x}_g$
\STATE $\mathbf{y}=\bigcup _{g \in \mathcal{G}}\mathbf{y}_g $
\end{algorithmic}
\end{algorithm}

To solve each subproblem $\mathcal{P}_g$, we use the method of Lagrange partial relaxation. Specifically, we relax the constraint (\ref{eq:dev_only_if_cache_con_gop}), which  is the ``hard'' constraint of the problem as it couples the caching and routing variables. As a result of relaxing this constraint, the Lagrangian subproblem consists of two optimization problems that can be solved independently: one that involves only the cache related variables of the original problem and another that involves only the delivery variables of the problem.

Specifically, let us define the set of Lagrange multipliers for every $g \in \mathcal{G}$ as follows:

\begin{align}
   \bm{\lambda}_g=( \lambda^{nu}_{vglt}\geq 0, &\forall n \in \mathcal{N},
    u \in \mathcal{U},  v \in \mathcal{V},  l \in \mathcal{L}, t \in \mathcal{T})
\end{align}


By relaxing the constraint in (\ref{eq:dev_only_if_cache_con_gop}) we obtain the Lagrangian function:
\begin{equation}
\begin{gathered}
\large
    L(\bm{\lambda}_g,\mathbf{x}_g,\mathbf{y}_g) = \\ \max_{\mathbf{x}_g,\mathbf{y}_g}   \sum_{n\in \mathcal{N}}\sum_{u\in \mathcal{U}}\sum_{v\in \mathcal{V}}\sum_{l\in \mathcal{L}}\sum_{t\in \mathcal{T}}(\frac{1}{\Delta} z^u_{vglt} \delta_{vglt} \alpha_{nu} - \lambda^{nu}_{vglt}) y^{nu}_{vglt}\\+
    \sum_{u\in \mathcal{U}}\sum_{v\in \mathcal{V}}\sum_{l\in \mathcal{L}}\sum_{t\in \mathcal{T}}\frac{1}{\Delta}z^u_{vglt} \delta_{vglt} y^{(N+1)u}_{vglt}\\+
     \sum_{n\in \mathcal{N} }\sum_{u\in \mathcal{U}}\sum_{v\in \mathcal{V}}\sum_{l\in \mathcal{L}}\sum_{t\in \mathcal{T}}\lambda^{nu}_{vglt} \alpha_{nu} x^{n}_{vglt}
\end{gathered}
\end{equation}

\noindent Thus, the Lagrangian dual problem can be expressed as:
\begin{equation}
\label{Lagrange_problem}
    \min_{\bm{\lambda}_g \geq \bm{0}}  L(\bm{\lambda}_g,\mathbf{x}_g,\mathbf{y}_g)
\end{equation}
where $\mathbf{x}_g,\mathbf{y}_g$ satisfy (\ref{eq:routing_var_gop})-(\ref{eq:delay_con_gop}), (\ref{eq:unsplittable_con_gop}), (\ref{eq:layer_con_gop}). We can easily observe that the Lagrange function can be re-expressed as follows:
\begin{equation}
    L(\bm{\lambda}_g,\mathbf{x}_g,\mathbf{y}_g)=f(\mathbf{x}_g)+k(\mathbf{y}_g)
\end{equation}
\noindent where $f(\mathbf{x}_g)$ and $k(\mathbf{y}_g)$ correspond to two independent optimization sub-subproblems $P1_g$ and $P2_g$, respectively. These sub-subproblems are defined as:

\begin{align}
& P1_g:\max_{\mathbf{x}_g}  \sum_{n\in \mathcal{N} }\sum_{u\in \mathcal{U}}\sum_{v\in \mathcal{V}}\sum_{l\in \mathcal{L}}\sum_{t\in \mathcal{T}}\lambda^{nu}_{vglt}  \alpha_{nu} x^{n}_{vglt}\label{eq:cachingcomp} \\ 
&   \textit{s.t.:}  & \nonumber\\
& \sum_{v \in \mathcal{V}} \sum_{l \in \mathcal{L}} \sum_{t \in \mathcal{T}} o_{vglt}  x^n_{vglt} \leq C_n(g), \forall{n \in \mathcal{N}}  \nonumber\\ 
&  \mathbf{x}_g = (x^{n}_{vglt} \in \{0,1\}:n \in \mathcal{N}, v \in \mathcal{V},  l \in \mathcal{L}, t \in \mathcal{T} )  \nonumber
\end{align}
%
\noindent and
\begin{align}
& P2_g:\max_{\mathbf{y}_g}  \sum_{n\in \mathcal{N}}\sum_{u\in \mathcal{U}}\sum_{v\in \mathcal{V}}\sum_{l\in \mathcal{L}}\sum_{t\in \mathcal{T}} ( \frac{1}{\Delta}z^u_{vglt} \delta_{vglt} \alpha_{nu} & \nonumber\\
&  -\lambda^{nu}_{vglt}) y^{nu}_{vglt} + \sum_{u\in \mathcal{U}}\sum_{v\in \mathcal{V}}\sum_{l\in \mathcal{L}}\sum_{t\in \mathcal{T}} \frac{1}{\Delta} z^u_{vglt} \delta_{vglt} y^{(N+1)u}_{vglt} \label{eq:routingcomp}&\\
& \textit{s.t.:} & \nonumber \\
& \sum_{n\in \mathcal{N_B}}y^{nu}_{vglt}\leq 1, \forall u\in \mathcal{U}, v\in \mathcal{V}, l\in \mathcal{L}, t\in \mathcal{T} & \nonumber \\
& \sum_{n \in \mathcal{N_B}}y^{nu}_{vg(l+1)t}\leq \sum_{n \in \mathcal{N_B}}y^{nu}_{vglt},  \forall   u\in \mathcal{U}, v\in \mathcal{V}, l\in \mathcal{L}, t\in \mathcal{T} & \nonumber \\
& \sum_{n \in \mathcal{N_B}}  \sum_{l \in \mathcal{L}} \sum_{t \in \mathcal{T}} o_{vglt}  d_{nu}  \alpha_{nu}  y^{nu}_{vglt} \leq t(g),  \forall u \in \mathcal{U}, v \in \mathcal{V} & \nonumber \\
& \mathbf{y}_g = (y^{nu}_{vglt} \in \{0,1\}: n \in \mathcal{N_B}, u \in \mathcal{U}, v \in \mathcal{V}, l \in \mathcal{L}, t \in \mathcal{T} ) & \nonumber 
\end{align}


Since $P1_g$ involves only the caching variables $\mathbf{x}_g$, we refer to it as the \textit{caching component}. In order to determine the optimal cache allocation $\mathbf{x}_g$ for $P1_g$, the problem is further decomposed into $N$ $0-1$ knapsack problems, which can be solved independently using optimization methods such as Dynamic Programming \cite{Knapsack_Problems}. In this case, we maximize the objective function of $P1_g$, considering that the cache space of each SBS is a knapsack. 

The optimization subproblem $P2_g$ involves only the routing decision variables $\mathbf{y}_g$, hence it is called hereafter as the \textit{routing component}. The subproblem $P2_g$ can be mapped to a multidimensional multiple choice knapsack problem. To find the optimal solution for the routing component we use CPLEX \cite{IBM}.

To solve the Lagrangian dual problem in (\ref{Lagrange_problem}), we apply the sub-gradient method to update the dual variables iteratively. Specifically, we start with non-negative values for the Lagrangian multiplier variables, and then based on the solution of the $P1_g$ and $P2_g$, in each iteration $\tau$ the dual variables are updated as follows:
\begin{equation}
    \lambda^{nu}_{vglt}(\tau+1)=[\lambda^{nu}_{vglt}(\tau)-\sigma(\tau) d(\lambda^{nu}_{vglt}(\tau))]^+ 
    \label{lamda_update}
\end{equation}

In (\ref{lamda_update}), $[s]^{+}=\max(0,s)$ is an operator that ensures that the dual variables cannot take negative values, and $\sigma(\tau)$ expresses the step size which controls the convergence properties of the sub-gradient algorithm at each iteration. In addition, $d(\lambda^{nu}_{vglt}(\tau))$ expresses the sub-gradient of the dual problem with respect of $\lambda^{nu}_{vglt}(\tau)$ and is given by:
\begin{equation}
   d(\lambda^{nu}_{vglt}(\tau))=-y^{nu}_{vglt}(\tau)+z^u_{vglt} \delta_{vglt} \alpha_{nu}  x^{n}_{vglt}(\tau)
\end{equation}

If we denote by $LB$ the Lower Bound of the subproblem $\mathcal{P}_g$, $UB$ the value of the Lagrange function at iteration $\tau$, and $\boldsymbol{\phi}(\tau) = d(\boldsymbol{\lambda}(\tau))$ the subgradient, the step size can be easily calculated using the formula below:
\begin{equation}
    \sigma(\tau)=w \frac{UB-LB}{\norm{\boldsymbol{\phi}(\tau)}^2}
\end{equation}

In the above equation, $w \in (0, 2]$ is a positive constant that scales the step size. The value of the Upper Bound is the value of the Lagrange function at each iteration, while the Lower Bound equals to the value of the objective function of the subproblem $\mathcal{P}_g$, by finding a feasible solution to it. The overall algorithm is summarized in Algorithm \ref{algo:CSlookup}, where the variable $\epsilon$ defines the maximum acceptable distance between the $UB$ and the $LB$. 

\begin{algorithm}[t]
\baselineskip=18pt
\caption{Primal-Dual Algorithm}
\label{algo:CSlookup}
\begin{algorithmic}[1]
\STATE \textbf{Require:} $\tau=1$, $\tau_{max}=1000$, $\boldsymbol{\lambda}_g(1)=0.2$, $UB=+\infty$, $LB=-\infty$, $w \in (0, 2]$, $\epsilon=0.01$
\WHILE {$\left|\frac{UB-LB}{LB}\right|\geq \epsilon$ \textbf{and} $\tau \leq \tau_{\max}$ }
\STATE Determine $\mathbf{x}_g(\tau)$ by solving $P1_g$
\STATE Determine $\mathbf{y}_g(\tau)$ by solving $P2_g$
\STATE $UB=L(\boldsymbol{\lambda_g},\mathbf{x}_g,\mathbf{y}_g)$ \text{and} $\sigma(\tau)=w \frac{UB-LB}{\norm{\boldsymbol{\phi}(\tau)}^2}$
\STATE Update Lower Bound (\textit{LB})
\STATE Update the dual variables $\bm{\lambda}_g(\tau+1)$ using (\ref{lamda_update})
\STATE Update $\tau=\tau+1$
\ENDWHILE
\end{algorithmic}
\end{algorithm}

\section{Performance Evaluation}
\label{sec:sim_results}
In this section, we evaluate the performance of the proposed algorithm exploiting cache collaboration and coding in multiple layers and tiles. First, we briefly describe all the schemes under comparison and give the simulation setup. Next, we provide experimental results that showcase the impact of the various system parameters on the performance of the proposed scheme. Finally, we discuss the convergence of the proposed algorithm.

\subsection{Simulation Setup}
\label{sec:simsetup}
The  schemes under comparison, including our proposed method, are described below:

\begin{enumerate}
\item \textit{Independent Caching-No Tiles (ICNT):} In this scheme, each video is encoded into a number of versions, where each version is encoded in a single tile and represents the whole scene in low quality along with a viewport in high quality. We assume that each user is associated only with one SBS, which has the maximum signal-to-interference-plus-noise ratio (SINR). The caching and delivery policy is found by solving (\ref{eq:Problem_formulation}) for each SBS separately. In this case, the caching and delivery variables are per video per GoP. 

\item \textit{Joint Caching-No Tiles (JCNT):} In this scheme, each video is encoded into a number of versions, where each version is encoded in a single  tile and represents the whole scene in low quality along with a viewport in high quality. Differently from ICNT, JCNT is a collaborative caching scheme that exploits the possible association of some users to more than one SBSs. The caching and routing decisions are jointly made using Algorithm 2. As previously, the caching and routing variables are per video per GoP as we do not use coding into multiple layers and tiles. 

\item \textit{Joint Caching-Multiple Layers (JCL):} In this scheme, each video is encoded in multiple quality layers and a single tile. In JCL, the base layer represents the whole scene in low quality, while the enhancement layers improve the quality of the part of the scene that corresponds to the viewport. The caching and routing policy is determined by solving Algorithm 2 on a per quality-layer basis. In this scheme, the caching and routing variables are per video per layer per GoP. 

\item \textit{Independent Caching (IC):} In this scheme, each video is encoded in multiple quality layers and tiles. Similar to ICNT we assume that each user is associated with only one SBS, i.e., the one with the maximum SINR. This scheme exploits the granularity of encoding the video into multiple tiles and quality layers. The caching and delivery policy is found by solving (\ref{eq:Problem_formulation}) for each SBS separately. In this case, the caching and delivery variables are per tile and layer per GoP. 

\item \textit{Proposed Algorithm:} This is the proposed scheme where videos are encoded in multiple quality layers and tiles. This scheme is similar to IC in that it takes advantage of the granularity of encoding the video into multiple tiles and quality layers. It further advances IC by exploiting collaboration opportunities among SBSs. In this scheme, the caching and routing decisions are jointly made on per tile and layer basis using Algorithm 2. 
\end{enumerate}

For both the proposed algorithm and JCNT the constant $\epsilon$ which controls the convergence is equal to 0.01. This ensures that UB and LB are close enough and the number of iterations required for the convergence of the algorithm is reasonable. The value of the weight $w$ is set to 0.02, which was found by experimentation. 

We assume for all conducted experiments a cellular network with 5 SBSs unless otherwise stated. The transmission radius of each SBS is set to $r_n=300$m, while each SBS has cache capacity sufficient to store 10\% of the size of the content library. The coverage radius of the MBS is set to $r_{N+1}=1000$m, which is large enough to allow communication between the MBS and all SBSs. We consider 30 users, who are randomly placed in the coverage area of the SBSs. The transmission delay from a SBS to a user is set to $d_{nu}=1$ sec/Mbit $\forall n \in \mathcal{N}, \forall u \in \mathcal{U}$, while the transmission delay when the content has to be fetched from the backhaul to the user is $d_{(N+1)u}=5$ sec/Mbit $\forall u \in \mathcal{U}$. 

The content library contains $V=10$ videos with resolution $1280 \times 720$. In our scheme the videos are encoded into $T=12$ tiles per frame and $L=2$ quality layers per tile. We assume that the size of each viewport is  $2 \times 2$ tiles. Each video consists of $G=30$ GoPs with duration of 1 sec per GoP. The playback delay $t_{app}$ for each video is 1sec, and the display time of a GoP $t_{disp}$ is 1 sec. The videos are encoded using the scalable extension (SHVC)\cite{SHVC_overview} of H.265/HEVC \cite{HEVC_description} standard, which allows encoding in tiles and layers.  Although the results we obtain are for video sequences encoded by SHVC, the derived conclusions are valid for videos encoded in tiles and layers by other codecs as well. Finally, we would like to note that we consider equirectangular projection, but our method is transparent to the employed projection, and hence is applicable when other projections such as cube or pyramid, etc. are used.

For the sake of simplicity, we assume that 40\% of the videos in the content library have similar characteristics with the ``Hog Rider'' video sequence, 30\% with the ``Roller Coaster'' video sequence, while the remaining 30\% with the ``Chariot Race'' video sequence. These videos were obtained from YouTube. The average size and the distortion reduction per tile for both base and enhancement layers are given in Table \ref{table:vid_parameters}.

\begin{table}
\centering
\captionsetup{font=scriptsize}
\caption{Distortion reduction per tile and layer for the considered video sequences}
\begin{tabular}{ |c|c|c|c|c| } 
\hline
\textbf{Video} & \boldsymbol{$o_{vg1t}$} & \boldsymbol{$o_{vg2t}$} & \boldsymbol{$\delta_{vg1t}$} & \boldsymbol{$\delta_{vg2t}$}\\
& (in Mbits) & (in Mbits) & & \\

\hline
\hline
``Hog Ride'' & 0.010 & 0.125 & 118 & 125  \\ 
\hline
``Roller Coaster'' & 0.208& 0.167 & 292 & 298 \\ 
\hline
``Chariot Racer'' & 0.029 & 0.275 & 187 & 192 \\ 
\hline
\end{tabular}
\label{table:vid_parameters}
\end{table}

\begin{figure}
    \begin{center}
     \includegraphics[width=9cm]{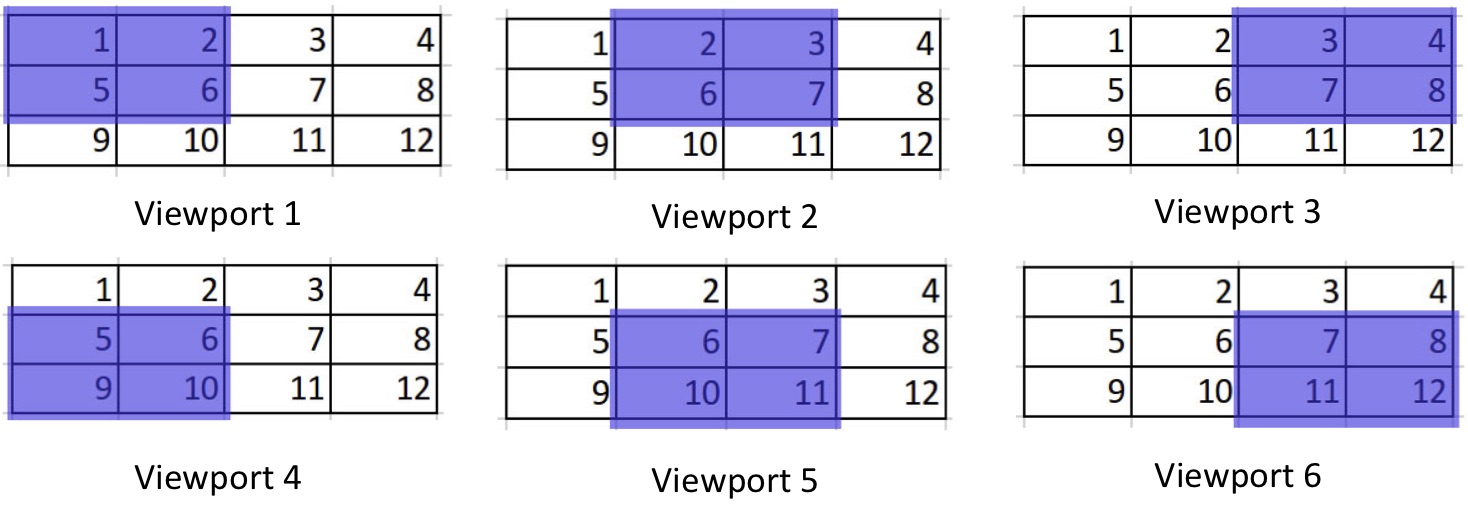}
     \end{center}
     \vspace{-0.5cm}
    \caption{Illustration of viewports considered for the evaluation. Numbers indicate tiles indices,  while highlighted areas denote the viewports.}
    \label{fig:possibleviewports}
\end{figure}

The probability that a specific video is requested by a user follows the Zipfian distribution \cite{zipf} with shape parameter $\eta=1$. Hence, the probability that a user requests the video $v \in \mathcal{V}$ is calculated as $$p_{v}=\frac{1/v^{\eta}}{ \sum_{v\in \mathcal{V} }1/ v^{\eta} }.$$ 
To study the effect of non-uniform popularity of tiles we consider that only the six viewports depicted in Fig. \ref{fig:possibleviewports} can be requested with non-zero probability by the users for each $360^o$ video. Assuming that the requests for these viewports are uniform, results in central tiles (which belong to multiple viewports) having higher popularity. We would like to emphasize that the set of viewports presented in Fig. \ref{fig:possibleviewports} is not complete; under equirectangular projection there are other available viewports that we do not consider here for the sake of obtaining the desirable distribution over the tiles and deriving interpretable results. This, however, does not change the conclusions drawn in this section.

As ICNT and JCNT schemes do not exploit encoding of $360^o$ video into tiles and layers, these schemes would be heavily penalized if a viewport were not cached in the SBSs caches as  data chunks of large size would not be delivered on time through the backhaul. Thus, for the sake of fairness, we allow soft satisfaction of users' demands \cite{Sermpezis18}  for these two schemes. Specifically, when a  viewport of a video requested by a user is not cached in an SBS, but another overlapping viewport of this video is cached, we assume that the cached viewport is delivered to the user. This way tiles that correspond to the overlap area of the requested and the delivered viewport are recovered at the highest quality, while the rest are recovered at the base quality. For example, if the requested and delivered viewports are viewports 1 and 2 in Fig. \ref{fig:possibleviewports}, respectively, tiles 2 and 6 will be recovered in high quality. In such a case, if the video is encoded in two quality layers, the soft cache hit ratio is computed by the following formula
\begin{equation}
\nonumber
    \text{SoftCHR} = \frac{nd_t^{b} + nd_t^{e}}{nr_t^{b} + nr_r^{e}} 
\end{equation}
where the numerator corresponds to the number of tiles delivered to the user and the denominator to the number of requested tiles. Specifically, $nd_t^{b}$ ($nd_t^{e}$) and $nr_t^{b}$ ($nr_t^{e}$) is the number of delivered tiles at base (enhancement) quality and requested number of tiles at base (enhancement) quality, respectively. For the example above, the numerical values are $nd_t^{b}=nr_t^{b}=12$, $nd_t^{e}=2$, and $nd_r^{e}=4$ and hence, the soft cache hit ratio is 0.875. We would like to note that soft satisfaction improves only the cache hit ratios of non-tiles based schemes (ICNT, JCNT, JLT). These schemes suffer from low cache hit ratio when the requested data cannot be delivered in time due to the use of larger chunks of data. In such case, the user can often obtain part of the required viewport through the delivery of another overlapping viewport. Through soft satisfaction cache hit ratio, we aim at capturing the improved interactivity a user experiences when instead of a required viewport, another viewport is delivered to the user. This happens as the user can navigate in part of the requested viewport but at degraded quality. Soft satisfaction assumes that tiles from base and enhancement layers are of equal importance and, hence, it is not a fully optimized QoE metric. The consideration of different weights for the importance of base and enhancement layers could lead to a more accurate QoE metric that better captures the tradeoff between quality and interactivity, but this is out of the scope of this paper.


\subsection{Parameter Analysis}

\subsubsection{Cache Size} 
\begin{figure}[t]
	\centering
		\includegraphics[width = 0.5 \textwidth]{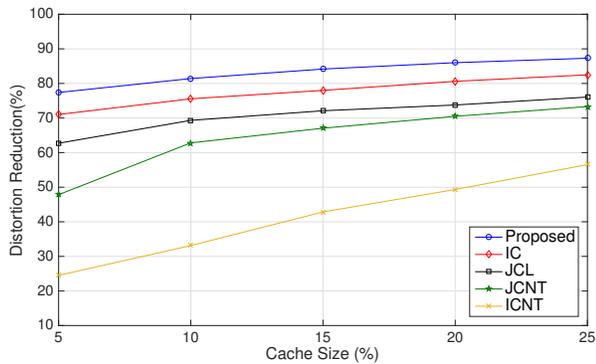}
		\caption{Distortion reduction with respect to the cache size for all schemes under comparison.}
	\label{fig:Comp_Cache}
\end{figure}

We first study the impact of the cache size of SBSs on the achieved distortion reduction $D$ as computed in (\ref{eq:objective}). We vary the cache capacity $C_n$ in the range [5,25]\% of the size of the content library and measure the distortion reduction as the percentage of the maximum achievable cumulative distortion reduction. The performance of the schemes under comparison is shown in Fig. \ref{fig:Comp_Cache}. From the results, we can see that the proposed scheme outperforms all the other schemes, for the entire range of SBSs' cache sizes. Specifically, we can observe that when SBSs can cache 5\% of the content library, which is common for networks handled by mobile network operators, the performance gap between the proposed algorithm and JCNT and ICNT, which do not assume encoding into multiple layers and tiles, is $\sim 30\%$ and $\sim 55\%$, respectively. This gap is due to the fact that the proposed algorithm takes advantage of the increased granularity in the caching and routing decisions offered by encoding the video in multiple tiles and layers, i.e., smaller chunks of data. This permits our algorithm to cache the most popular and significant chunks of data locally in the SBSs, while preserving high data diversity across the SBSs thanks to collaboration among SBSs. Similar trends but smaller gains, $\sim 15\%$, can be observed when we compare the proposed algorithm with JCL, which assumes encoding into quality layers and a single tile. The observed gains are because the proposed algorithm benefits from the encoding in tiles, unlike JCL which only considers layers. Specifically, in the proposed scheme the cached tiles are obtained locally from the SBS, while the rest of the tiles are delivered from the backhaul within the delivery deadlines. This is often not possible when JCL is used, as the delay constraint may not allow timely delivery of the larger data chunks that correspond to quality layers. Finally, to understand the gains arising from collaborative caching we compare the proposed scheme with IC. We note that the proposed algorithm still outperforms the IC scheme despite the fact that the latter uses tile-based and layered encoding. This performance difference is attributed to the exploitation of the collaborative caching opportunities among SBSs which leads to gains when users reside in the overlap area between the coverage area of multiple SBSs. As we will show later in this section, the gains grow when the coverage area of the SBSs increases, i.e. more users are in the overlap area of several SBSs. We will also show that SBSs collaboration is associated with even higher cache hit ratio, i.e., less data has to be fetched through the backhaul.
 
As we can further observe from Fig. \ref{fig:Comp_Cache} when SBSs' cache capacity increases, the performance gap between the proposed algorithm and the under comparison schemes in most of the cases becomes smaller. Specifically, when each SBS can cache $25\%$ of the content catalogue, the proposed algorithm outperforms JCNT and ICNT by $\sim 15\%$ and $\sim 30\%$, respectively. The performance gap is smaller for such cache sizes because larger cache capacity permits ICNT and JCNT to cache more videos in SBSs and less requests need to be served through the backhaul links, enabling the timely delivery of more data. The performance gap between the proposed algorithm and the JCL is $\sim 10\%$. JCL cannot close further the gap, as the data chunks corresponding to the layers are of larger size. When we compare the proposed scheme with IC we note that the performance gap stays constant because both schemes takes advantage of the additional cache size but the proposed scheme in addition exploits collaboration opportunities among SBSs. However, it is expected that both schemes will perform identical if the cache size is further increased, as most of the data will be cached locally and the backhaul link will not be used. These results are not presented as they are of no practical interest.

\begin{figure}[t]
	\centering
		\includegraphics[width = 0.5 \textwidth]{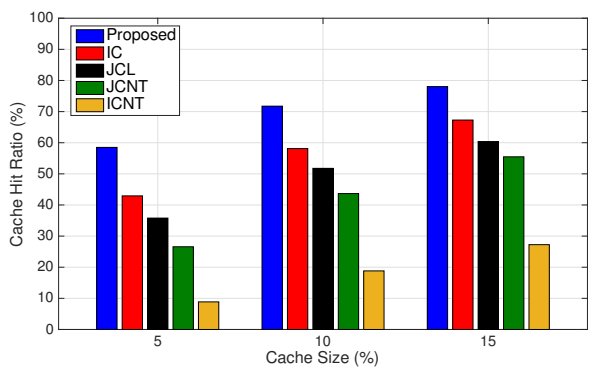}
		\caption{Cache hit ratio with respect to the cache size for all schemes under comparison.}
		\vspace{-0.5cm}
	\label{fig:Cache_hit_ratio}
\end{figure}

The above discussion regarding the benefits of making joint routing and caching decisions on per layer and tile basis are verified by Fig. \ref{fig:Cache_hit_ratio}, which depicts the achieved cache hit ratio for various cache sizes for all the schemes under comparison. From this figure, we can note that for all cache sizes the proposed scheme achieves higher cache hit ratio. This means that the majority of requests are served locally from the SBSs caches without the need to use the backhaul links. On the contrary, the low cache hit ratio in the ICNT and JCNT schemes indicates that user requests are not served locally and content is fetched from the remote servers, which due to increased delivery delay and chunk size (compared to tile size in our scheme) may not be always delivered within the time constraints. This, in turn, results in a small distortion reduction. When we compare the proposed scheme with JCL and IC we can observe that for 10\% cache size, it outperforms these schemes by $\sim 20\%$ and $\sim 14\%$, respectively. The performance difference is significant, however it is smaller than that between the proposed scheme and ICNT and JCNT. This is because JCL and IC encode the data in smaller chunks than ICNT and JCNT, i.e., JCL considers encoding in layers and IC encoding in layers and tiles. The gains of the proposed scheme over IC are due to the exploitation of collaborative caching opportunities.

\subsubsection{SBS radius} 
\begin{figure}[t]
	\centering
		\includegraphics[width = 0.5 \textwidth]{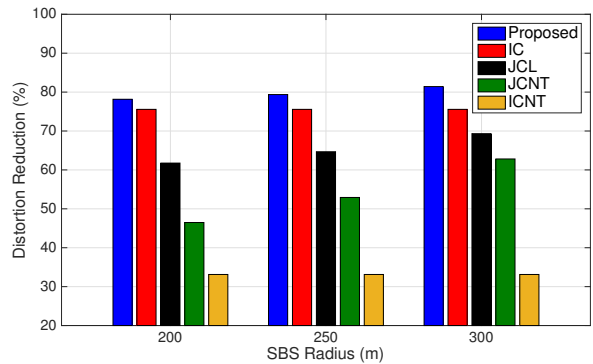}
		\caption{Distortion reduction with respect to the radius of the SBSs for all schemes under comparison.}
	\label{fig:Comp_Radius}
\end{figure}

To investigate the impact of the collaboration we change the transmission radius (coverage area) of the SBSs on the distortion reduction from 200m up to 300m. The results are depicted in Fig. \ref{fig:Comp_Radius} from where we can note that an increase in the SBS radius affects the performance of the schemes performing collaborative caching, i.e., the proposed algorithm, JCL and the JCNT scheme, while the performance of non-collaborative caching schemes, i.e., ICNT and IC, stays invariant. This can be explained by the fact that for the collaborative caching schemes a higher overlap in the coverage area of the SBSs results in more users being in the overlap area of multiple SBSs. On the contrary, for the non-collaborative caching schemes an increase in the coverage area is not translated into performance gains because each user is associated with a single the SBS, that with the maximum SINR. From the results, we can further see that the proposed scheme in all cases outperforms all other schemes. Additionally, the performance gap between the proposed scheme and IC grows from $\sim 2\%$ to $\sim 6\%$, which makes clear the significance of exploiting SBSs collaboration opportunities. The scheme that profits the most from the increase in SBS transmission radius is JCNT which considers the largest chunks of data (entire videos) and for this scheme, cache collaboration partially compensates for the size of the data chunks.

\subsubsection{SBS communication link delay} 
\begin{figure}[t]
	\centering
		\includegraphics[width = 0.5 \textwidth]{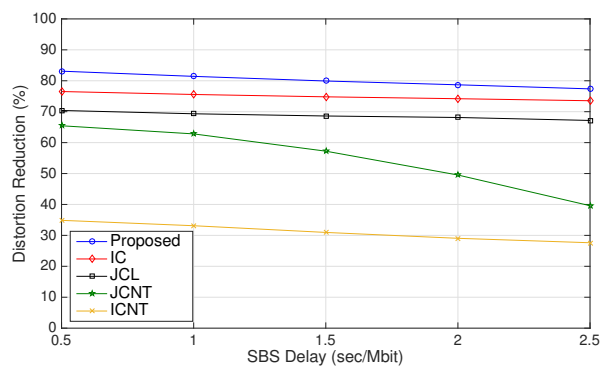}
		\caption{Distortion reduction with respect to the SBS Delay for all schemes under comparison.}
		\vspace{-0.5cm}
	\label{fig:SBS_Delay}
\end{figure}

In Fig. \ref{fig:SBS_Delay}, we illustrate the effect of the delay of the communication link between the users and the SBSs on the distortion reduction $D$. In particular, we assume that the value of the SBS delay $d_{nu}$ varies in the range $[0.5, 2.5]$ sec/Mbit, while the backhaul delay remains constant at 5 sec/Mbit. We can observe that an increase in the SBS delay (i.e., the communication link becomes slower) results in a decrease in distortion reduction $D$ for all the schemes. We can note that the performance of the proposed algorithm is only slightly affected by the increase in the SBS delay ($\sim 5\%$). Further, we can see that IC behaves similar to the proposed algorithm. This behavior can be explained by the fact that both schemes exploit tiles encoding; the superior performance of the proposed scheme is due to the exploitation of the collaboration opportunities between SBSs. We can see that  JCNT and ICNT are more affected by the higher SBS communication link delays.  As a result, the performance difference between the proposed algorithm and ICNT and JCNT grows from $\sim 48\%$ to $\sim 50\%$ and from $\sim 18\%$ to $\sim 28\%$, respectively, as the link delay increases from 0.5 to 2.5 sec/Mbit. This happens as ICNT and JCNT schemes that do not employ encoding into tiles and layers are affected the most by the increase in the SBSs link delay since this prevents the timely delivery of large data chunks, resulting in a significant decrease in the achieved distortion reduction. JCL, which considers encoding in layers, performs better than ICNT and JCNT because the employed data chunks are of smaller size, which means that they are affected less from an increase in the SBSs communication link delay. From this comparison, it is evident that caching of smaller chunks of data is beneficial and that collaborative caching and delivery decisions can bring additional gains.

\subsubsection{Backhaul link delay} 
\begin{figure}[t]
	\centering
		\includegraphics[width = 0.5 \textwidth]{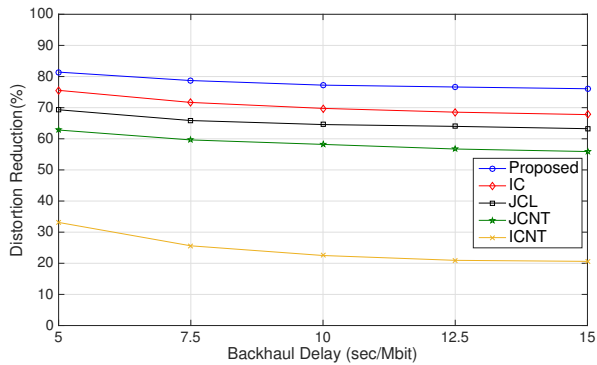}
		\caption{Distortion reduction with respect to the Backhaul Delay for all schemes under comparison.}
	\label{fig:Comp_Backhaul}
\end{figure}

In Fig. \ref{fig:Comp_Backhaul}, we compare the performance of all the schemes with respect to the backhaul delay. Specifically, we consider that the backhaul delay $d_{(N+1)u}$ takes values in the range [5, 15] sec/Mbit. From the results, we can conclude that an increase in the backhaul link delay results in users experiencing a lower distortion reduction. This happens because an increase in the backhaul delay results in less data being delivered through the backhaul link within the time constraint. The increase in the backhaul delay has a more significant impact on the JCNT and ICNT, while there is a gentle decrease for all other schemes. In particular, the difference in the distortion reduction between the proposed method and JCL stays above 12\%, while the difference between the proposed method and the IC varies from $\sim 6\%$ to $\sim 9\%$. This is because of JCL encoding into layers, and the proposed scheme and IC encoding in both tiles and layers. By increasing the backhaul delay, the proposed scheme and IC are not affected significantly as the tiles (layers for JCL) of the requested videos are stored and delivered by the SBSs. Also, due to the smaller size of the tiles (layers for JCL) compared to videos, more tiles can be delivered in time to the users through the backhaul even when the backhaul delay increases. ICNT and JCNT are affected more by this delay increment, and the performance gap grows  from $\sim 48\%$ and $\sim 18\%$  to $\sim 56\%$ and $\sim 20\%$, respectively, for a delay of 5 sec/Mbit.

\subsubsection{Video popularity distribution} 
\begin{figure}[t]
	\centering
		\includegraphics[width = 0.5 \textwidth]{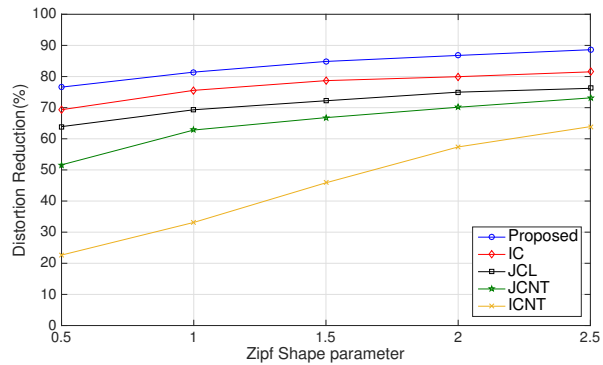}
		\caption{Distortion reduction with respect to Zipf shape parameter for all schemes under comparison.}
		\vspace{-0.5cm}
	\label{fig:Comp_Zipf}
\end{figure}

In Fig. \ref{fig:Comp_Zipf}, we examine the effect of the skewness parameter of the Zipfian distribution on the distortion reduction. To this end, we vary the shape parameter $\eta$ from $0.5$ up to $2.5$. For small values of $\eta$, i.e., very diverse content demands, the difference between the proposed scheme and ICNT and JCNT is approximately 54\% and 25\%, respectively. This difference is due to the encoding in tiles which permits to cache the base layer tiles locally and hence satisfy most of the users' requests with a basic video quality. This is not the case in ICNT and JCNT as without encoding in tiles and layers the local caches are occupied by large data chunks which can satisfy only a small percentage of the diverse users' requests. Similar to previous comparisons we can see that performance difference between the proposed algorithm and JCL and IC is smaller than that with ICNT and JCNT because former schemes consider data chunks of smaller size, layers for JCL and tiles for IC. The proposed scheme outperforms IC because of the exploitation of cache collaboration opportunities. The performance gap closes when $\eta$ value becomes higher, as the majority of the user requests refer to a smaller number of videos. Higher values of $\eta$ are more beneficial for ICNT and JCNT, as these schemes do not encode videos in tiles and layers, but even for such $\eta$ values, ICNT and JCNT schemes performance remains inferior to that of the other schemes. It is worth to note that for higher $\eta$ values collaboration becomes less important, as there is less variety in the content requests. This becomes clear when we compare the trend of ICNT and IC with JCNT and the proposed scheme.

\color{black}

\subsubsection{Viewports popularity distribution}
\begin{figure}[t]
	\centering
		\includegraphics[width = 0.5 \textwidth]{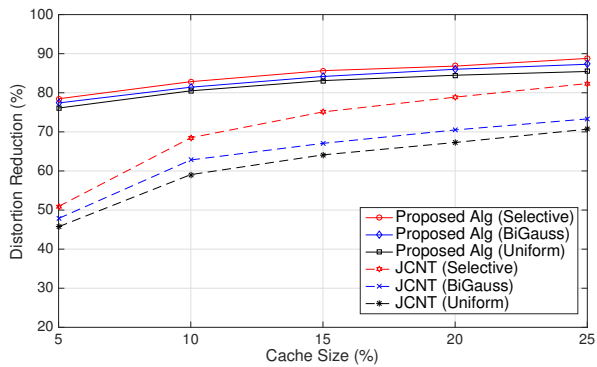}
		\caption{Distortion reduction with respect to the cache size for the proposed scheme and JCNT considering three viewport popularity distributions.}
	\label{fig:Comp_ViewportPop}
\end{figure}

We also study the effect of viewports popularity on the performance of the considered schemes for SBSs' radius equal to 200m. To this aim, we examine three viewport popularity distributions: (a) the viewport popularity distribution described in Section \ref{sec:simsetup}, which we term hereafter as ``BiGauss'' as according to this distribution the central tiles are more popular than the rest; (b) a uniform distribution where we consider all possible viewports (allowing viewports to fold around) with uniform popularity which results in uniform popularity of tiles; and (c) a selective viewport distribution where for each video all users request the same viewport of this video. We compare the performance of the proposed scheme and the JCNT scheme.  The results are presented in Fig. \ref{fig:Comp_ViewportPop}. As expected the more concentrated are users' requests in fewer tiles, the higher is the distortion reduction for both schemes. Further, we can see that the proposed scheme outperforms significantly JCNT because caching and delivery decisions are made per tile and tiles are of small size which  means they can be delivered to the users within the delivery deadlines. The performance gains become smaller when the cache size increases. This is attributed to the fact that when cache space increases, JCNT can cache more than one viewports of popular videos assuming BiGauss or uniform distribution and more videos for selective distribution. For the proposed scheme the distortion reduction improvements are smaller, as already for small cache sizes our scheme is able to achieve distortion reductions that exceeds 75\%. In the proposed scheme even when cache resources are scarce, thanks to tiles and layers granularity, the most popular tiles are cached locally at the SBS and the rest can be delivered through MBS.

\subsection{Convergence}
\begin{figure}[t]
	\centering
		\includegraphics[width = 0.5 \textwidth]{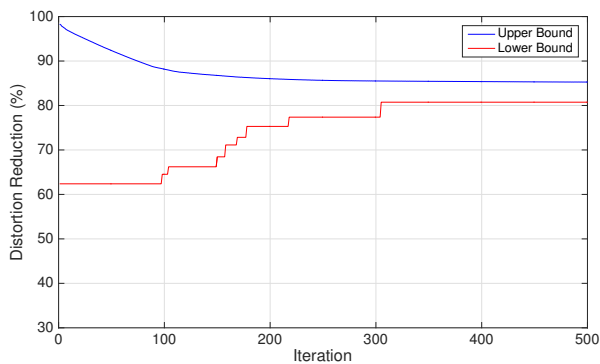}
		\caption{Convergence of the proposed algorithm.}
	\label{fig:Convergence}
\end{figure}

For the sake of completeness, we examine the convergence of the proposed algorithm. Recall that each subproblem $\mathcal{P}_g$ involves routing and caching decisions of $g$th GoP and is obtained by solving iteratively Algorithm 2. The convergence for subproblem $\mathcal{P}_1$, i.e., for the first GoP, is depicted in Fig. \ref{fig:Convergence}. Similar convergence behavior is noticed for all GoPs, but is omitted here for brevity reasons. From Fig. \ref{fig:Convergence}, we can see that the proposed algorithm requires only a few hundreds iterations to converge. This is in accordance with the work in \cite{Subgradient_Poularakis}, where it was shown that the Lagrange partial relaxation method provides an upper bound when applied to $\mathcal{P}_g$, which is guaranteed to converge to the optimal value of the $\mathcal{P}_g$ in a finite number of iterations. 


\section{Conclusion}
\label{sec:concl}
In this work, we studied the problem of the $360^o$ video delivery and caching in cellular networks comprising an MBS and multiple SBSs that can collaborate. To maximize the quality of the video delivered to the end users, we exploit advanced video coding tools offered by video coding standards such as HEVC that permits encoding of video in multiple tiles and layers offering greater granularity of information. We also exploit SBSs collaboration opportunities to prevent neighboring SBSs from caching the same video data which reduces the load of the backhaul link. The proposed algorithm takes into consideration not only the content importance but also video popularity at tile level, to jointly decide the routing and caching policies. As the original problem is of high complexity, we decompose it into a number of subproblems, one per each GoPs, which are then solved sequentially. To further reduce the complexity of the proposed algorithm, we decouple the subproblems into their routing and caching components and solve them using the Lagrange decomposition method. The experimental evaluation shows that collaborative caching and video encoding into tiles and quality layers allows to cache the most important data locally in the SBSs and deliver it in a timely manner to the users. Hence, the system outperforms significantly its counterparts that do not use collaboration and/or encoding into multiple tiles and quality layers. We would like to close with the remark that our method does not preclude the use of a predictive mechanism during the actual video delivery. That is, once the optimal cache allocation has been done and the video is streamed to the user, if a prediction mechanism is in place, it can be used to further enhance the experience of the user.

\bibliographystyle{IEEEtran}

\end{document}